\documentclass[12pt]{article}
\usepackage{graphicx}

\newcommand\pubnumber{}
\newcommand\pubdate{}

\def\napoli{Institute for Nuclear Physics\\
Johannes Gutenberg-University Mainz, 55128 Mainz, GERMANY}
\def\support{}

\def\Title#1{\begin{center} {\Large #1 } \end{center}}
\def\Author#1{\begin{center}{ \sc #1} \end{center}}
\def\Address#1{\begin{center}{ \it #1} \end{center}}

\newcommand\pubblock{\rightline{\begin{tabular}{l} \pubnumber\\
         \pubdate  \end{tabular}}}
\newenvironment{Abstract}{\begin{quotation}  }{\end{quotation}}
\newenvironment{Presented}{\begin{quotation} \begin{center} 
             PRESENTED AT\end{center}\bigskip 
      \begin{center}\begin{large}}{\end{large}\end{center} \end{quotation}}

\def\beq{\begin{equation}}
\def\eeq#1{\label{#1}\end{equation}}
\def\eeqn{\end{equation}}

\def\beqa{\begin{eqnarray}}
\def\eeqa#1{\label{#1}\end{eqnarray}}
\def\eeqan{\end{eqnarray}}

\let\bar=\overbar

\def\Dslash{\not{\hbox{\kern-4pt $D$}}}
\def\dslash{\not{\hbox{\kern-2pt $\del$}}}

\def\msb{{\bar{\ssstyle M \kern -1pt S}}}

\begin{document}
\begin{titlepage}
\pubblock

\vfill
\Title{Measurement of Hadronic Cross Sections at BESIII}
\vfill
\Author{ Christoph Florian Redmer for the BESIII Collaboration\support}
\Address{\napoli}
\vfill
\begin{Abstract}
The uncertainties of the Standard Model prediction of the anomalous magnetic moment of the muon are currently completely 
dominated by hadronic contributions. The largest contribution is due to the hadronic vacuum polarization. Hadronic cross 
sections measured at $e^+e^-$ colliders can be exploited as experimental input to improve the calculations, making use 
of the optical theorem. At the BESIII experiment in Beijing these cross sections are determined using different 
methods. At center-of-mass energies above 2 GeV exclusive and inclusive cross sections can be measured in an energy 
scan. Additionally, cross sections can be determined starting from the $\pi^+\pi^-$ mass threshold using the method of 
Initial State Radiation. An overview of the recent results and the status of the analyses is provided.
\end{Abstract}
\vfill
\begin{Presented}
Thirteenth International Conference on the Intersections of Particle and Nuclear Physics (CIPANP2018)\\
Palm Springs, California, May 28--June 3, 2018
\end{Presented}
\vfill
\end{titlepage}
\def\thefootnote{\fnsymbol{footnote}}
\setcounter{footnote}{0}

\section{Anomalous magnetic moment of the muon}
The muon anomaly $a_\mu=\frac{g_\mu - 2}{2}$ describes the relative deviation of the muon's gyromagnetic factor from 
the value expected in Dirac theory, which is $g_\mu=2$. It is one of the most precisely determined parameters in the 
Standard Model (SM). The theory prediction as well as the direct measurement have achieved an accuracy on the level of 
0.5~ppm. However, there is a long standing discrepancy between experiment and theory, which differ by more than three 
standard deviations~\cite{Blum:2013xva}. This difference gives rise to numerous activities in experiment as well as 
theory in order to understand if it should be considered as a hint for New Physics.

The most recent direct measurement of $a_\mu$ was performed by E821 at BNL~\cite{Bennett:2006fi}. Two new and 
systematically independent direct measurements of $a_\mu$ are planned. Both aim at a fourfold improvement of the 
Brookhaven result. The E989 experiment at Fermilab reuses the BNL storage ring~\cite{Grange:2015fou}. Higher beam 
intensities and an improved apparatus are used to bring down the errors. It is expected to reproduce the accuracy of the 
BNL result by spring 2019. The second experiment is under construction at J-PARC~\cite{Mibe:2011zz}. A new and 
independent approach is applied by using a beam of ultra-cold muons, which allows to avoid the use of focusing electric 
fields.

The SM prediction of $a_\mu$ takes into account contributions from electromagnetic, weak and strong interactions. While 
the first two are well understood and known with good accuracy, the latter completely dominates the uncertainty of the 
SM prediction. Due to the running of the strong coupling constant, the contribution cannot be treated in perturbation 
theory at the relevant energies. The contribution of strong interaction is separated into two parts. On the one hand 
there is the contributions due to the hadronic vacuum polarization $a_\mu^{hVP}$, and on the other hand there is the 
contribution due to the hadronic light-by-light scattering $a_\mu^{hLbL}$. Recently, there have been a lot of activities 
to calculate these contributions in lattice QCD~\cite{lqcd}. Another approach is to use experimental data as input to 
the calculations for the SM prediction. Especially the leading order contribution of $a_\mu^{hVP}$ can be systematically 
improved in this way. Here, the optical theorem relates the vacuum polarization to hadronic cross sections, which can 
be measured in $e^+e^-$ annihilation. By increasing the accuracy of the cross section measurements, the uncertainty of 
$a_\mu^{hVP}$ is reduced. The measured cross sections are evaluated in a dispersion integral, which also contains a 
kernel function. Both the kernel function as well as the cross sections show an energy dependence, which decreases with 
the square of the center-of-mass energy. Thus, the knowledge of hadronic cross sections at $\sqrt{s}<1\,\textrm{GeV}$ 
is of utmost importance. This energy range is dominated by the $\rho$ resonance, which predominantly decay into 
$\pi^+\pi^-$ pairs. The cross section $\sigma(e^+e^-\to\pi^+\pi^-)$ makes up for more than 70\% of the value of 
$a_\mu^{hVP}$. It is also the dominating contribution to the uncertainty of $a_\mu^{hVP}$, however, here also higher 
pion multiplicities, and final states with kaons play a significant role. In order to contribute to the efforts to 
improve the SM prediction of $a_\mu$, the BESIII collaboration has started a program to measure hadronic cross sections 
with high accuracy.

\section{The BESIII detector}
The BESIII detector is a magnetic spectrometer~\cite{Ablikim:2009aa} located at the Beijing Electron Positron Collider 
(BEPCII)~\cite{Yu:IPAC2016-TUYA01}. The cylindrical core of the BESIII detector consists of a helium-based  multilayer 
drift chamber (MDC), a plastic scintillator time-of-flight system (TOF), and a CsI(Tl) electromagnetic calorimeter 
(EMC), which are all enclosed in a superconducting solenoidal magnet providing a 1.0~T magnetic field. The solenoid is 
supported by an octagonal flux-return yoke with resistive plate counter muon identifier modules interleaved with steel. 
The acceptance of charged particles and photons is 93\% over $4\pi$ solid angle. The charged-particle momentum  
resolution at $1~{\rm GeV}/c$ is $0.5\%$, and the $dE/dx$ resolution is $6\%$ for the electrons from Bhabha scattering. 
The EMC measures photon energies with a resolution of $2.5\%$ ($5\%$) at $1$~GeV in the barrel (end cap) region. The 
time resolution of the TOF barrel part is 68~ps, while that of the end cap part is 110~ps.

The accelerator BEPCII provides $e^+e^-$ collisions at center-of-mass energies between $\sqrt{s}=2.0\,GeV$ and
4.6\,GeV. The performance in terms of peak luminosity is optimized for data taking at $\sqrt{s}=3.773\,\textrm{GeV}$, 
which corresponds to the peak of the $\psi(3770)$ resonance. The design luminosity of $10^{33}\,\textrm{cm}^{-2} 
\textrm{s}^{-1}$ is reached. Over the past years large data samples have been collected, which are used to pursue the 
BESIII physics program, focusing on charm physics, charmonium and charmoniumlike spectroscopy, light hadron physics, 
QCD tests, and precise $\tau$ mass measurements.

Hadronic cross section are measured at BESIII using different techniques. In the energy range covered by BEPCII the 
energy dependence of cross sections can be studied in a conventional energy scan. It is used to investigate exclusive 
as well as inclusive cross section, like the R-ratio, which relates the total inclusive hadronic cross section to the 
cross section of muon pair production.

Recently, also the method of initial state radiation (ISR) is used to extend the accessible energy range. The emission 
of 
a photon from the initial state lowers the effective center-of-mass energy $\sqrt{s^\prime} = 
\sqrt{s-2\sqrt{s}E_\gamma}$, where $E_\gamma$ is the energy of the ISR photon. The method allows to study hadronic 
cross sections down to $\sqrt{s}=2m_\pi$. However, cross sections measured through ISR events are radiative cross 
sections. The cross section of the non-radiative process can be calculated, taking into account the radiator function 
$H(s, E_\gamma,\theta_\gamma)$, which describes the probability at a specific $\sqrt{s}$ to emit an ISR photon of the 
energy $E_\gamma$ at the polar angle $\theta_\gamma$. The relation of radiative and non-radiative cross sections is 
given by $\frac{d\sigma_{\rm had + \gamma}}{dm_\gamma} = \frac{2 m_{\rm had}}{s} H(s, E_\gamma, \theta_\gamma) 
\sigma_{\rm had}$.

The ISR photons are emitted with a characteristic angular distribution, where the emission along the initial lepton 
beam direction is strongly preferred. Due to momentum conservation, a hadronic system produced at $\sqrt{s^\prime}$ is 
boosted to the opposite direction. Depending on the possibility to register the ISR photon in the BESIII detector, two 
different analysis strategies are defined for ISR events. When the hadronic system as well as the ISR photon are 
detected the measurement is referred to as ``tagged ISR measurement''. In this case, cross sections can be measured from 
$\sqrt{s^\prime}=2m_\pi$ in principle up to $\sqrt{s}$. However, the larger the mass of the produced hadronic system, 
i.e. $\sqrt{s^\prime}$ becomes, the higher is the background contamination. Random signals in the EMC, e.g. from noise 
or machine background, can easily be mistaken the low energetic ISR photon.
This is much less of a problem when the ISR photon escapes detection by being emitted along the beam pipe and only the 
hadronic system is measured. This type of measurement is referred to as ``untagged ISR measurement''. The four-momentum 
of the unmeasured photon can be reconstructed using energy and momentum conservation. By rejecting event candidates, 
where the momentum does not point along the beam axis, background is rejected efficiently. At the same time this 
analysis strategy selects the peaking part of the differential cross section of the ISR process, allowing for a high 
statistics measurement. A slight restriction in the applicability of the method comes from the boost of the hadronic 
system. The higher the energy of the emitted photon, the more the hadrons are boosted towards the beam axis, where the 
detector acceptance is limited due to the accelerator. Thus, there is a lower limit for the mass of hadronic systems to 
be measured in the untagged ISR method. The acceptance of the BESIII detector imposes a threshold of approximately 
1\,GeV/c$^2$.

The radiation of a photon from the initial state is a higher order effect. Radiative processes are suppressed by a 
factor $\frac{\alpha}{\pi}$. Precision studies performed using either of the ISR methods require large data samples. The 
BESIII collaboration has acquired data sets of more than $10\,\textrm{fb}^{-1}$ at $\sqrt{s} \geq 
3.773\,\textrm{GeV}$~\cite{Ablikim:2013,Ablikim:2015nan}. Hadronic cross sections are measured based on these data 
sets making use of both the tagged and untagged ISR methods.

\section{\boldmath $e^+e^-\to\pi^+\pi^-$}
Since the cross section $\sigma(e^+e^-\to\pi^+\pi^-)$ make up for more than 70\% of $a_\mu^{hVP}$, it is of utmost 
importance to have data with highest accuracy to improve the SM prediction. The pion production process has been 
studied 
by several experiments in the past. The results with leading accuracy come from the KLOE~\cite{Anastasi:2017eio} and 
BaBar~\cite{Aubert:2009ad} collaborations. Both claim sub-percent accuracy for their results. However, the data differ 
by more than 3\%, which is also reflected in the uncertainty of the evaluation of $a_\mu^{hVP,LO}$. In order to clarify 
the situation, a new high precision measurement has been performed at BESIII~\cite{Ablikim:2015orh}. Using the tagged 
ISR method, a data set of 2.93\,fb$^{-1}$ taken at $\sqrt{s}=3.773\,\textrm{GeV}$ has been analyzed. The cross section 
of two-pion production at $600 \leq \sqrt{s}[\,\textrm{GeV}]\leq 900$ is investigated selecting the final state 
$\pi^+\pi^-\gamma$. Radiative production of muon pairs in $e^+e^-\to\mu^+\mu^-\gamma$ is the dominating background 
contribution. The similarity of the signals of pions and muons in most detectors make it hard to distinguish them based 
on a single source of information. An artificial neural network (ANN) has been designed combining the information from 
different sub-detectors, like energy loss in the MDC, energy deposits and shower shapes in the EMC, and the penetration 
depth of a particle in the MUC. Taking carefully into account systematic differences between data and simulation, the 
ANN is trained with Monte Carlo samples. It allows  to effectively separate muons from pions. The correct operation of 
the ANN and the validity of systematic corrections has been tested by comparing the muon event yield in data with the 
QED prediction as implemented in the \textsc{Phokhara} event generator~\cite{phokhara}, where an accuracy of 0.5\% is 
claimed. Excellent agreement with the selected data is found.

Finally, the $e^+e^-\to\pi^+\pi^-\gamma_{\rm ISR}$ signal event yield is corrected for detection efficiency, and is 
normalized to the integrated luminosity of the data set to obtain the radiative cross section. By correcting for vacuum 
polarization and FSR effects, and by dividing out the radiator function, the bare cross section, relevant to determine 
the contribution to $a_\mu^{hVP,LO}$, is calculated. A total uncertainty of the cross section of 0.9\% is achieved. The 
dominating systematic uncertainties are due to the luminosity determination and the knowledge of the radiator function, 
with 0.5\% each. Extracting the cross section by normalizing to the muon yield might avoid these two contributions, 
however, the statistical uncertainty of the muon yield in the analyzed data does not allow for a result with the aimed 
accuracy.

\begin{figure}[htb]
 \centerline{
  \includegraphics[trim=0mm 0mm 9mm 5mm, clip, height=4cm]{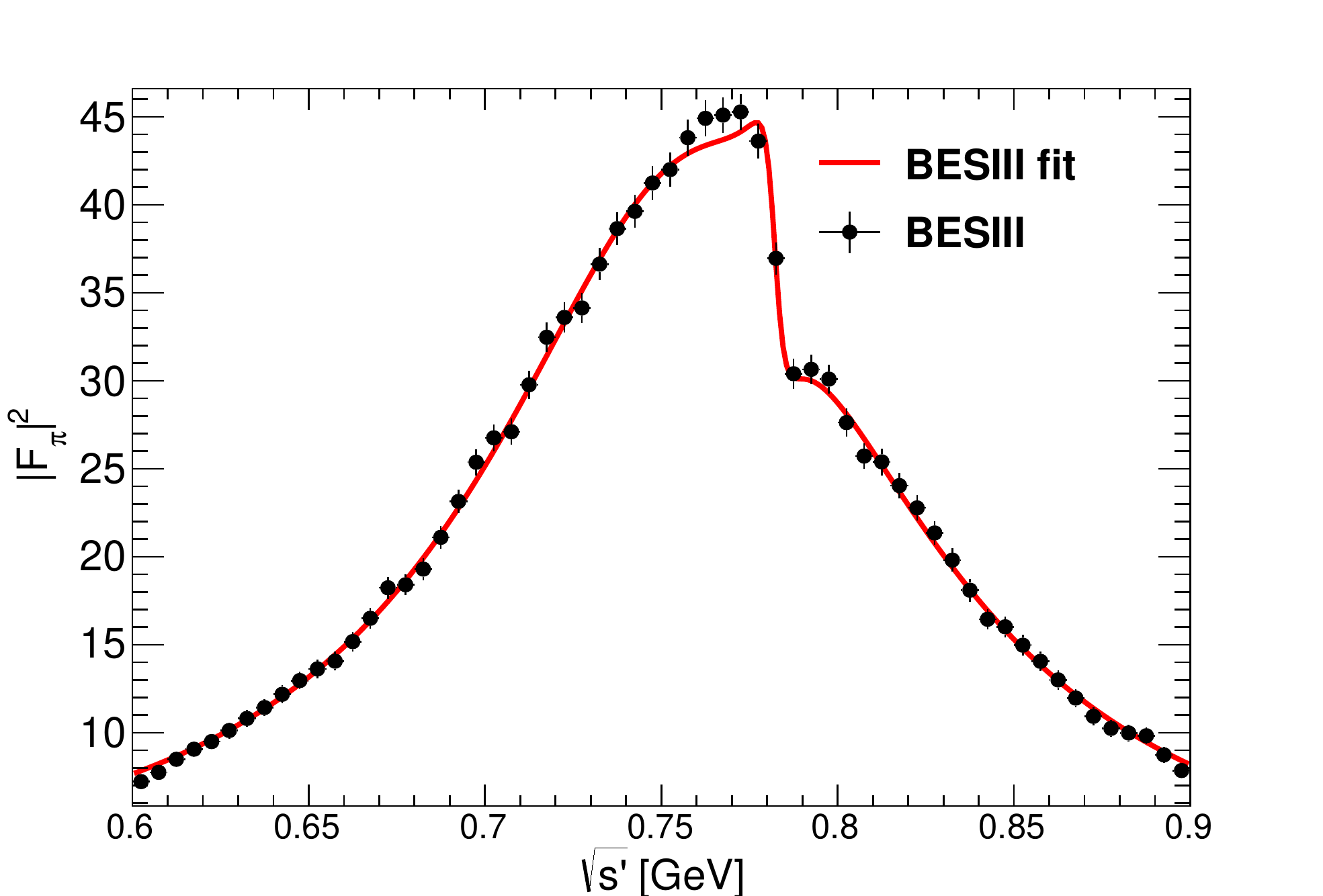}\hfill%
  \includegraphics[trim=13mm 0mm 18mm 5mm, clip, height=4cm]{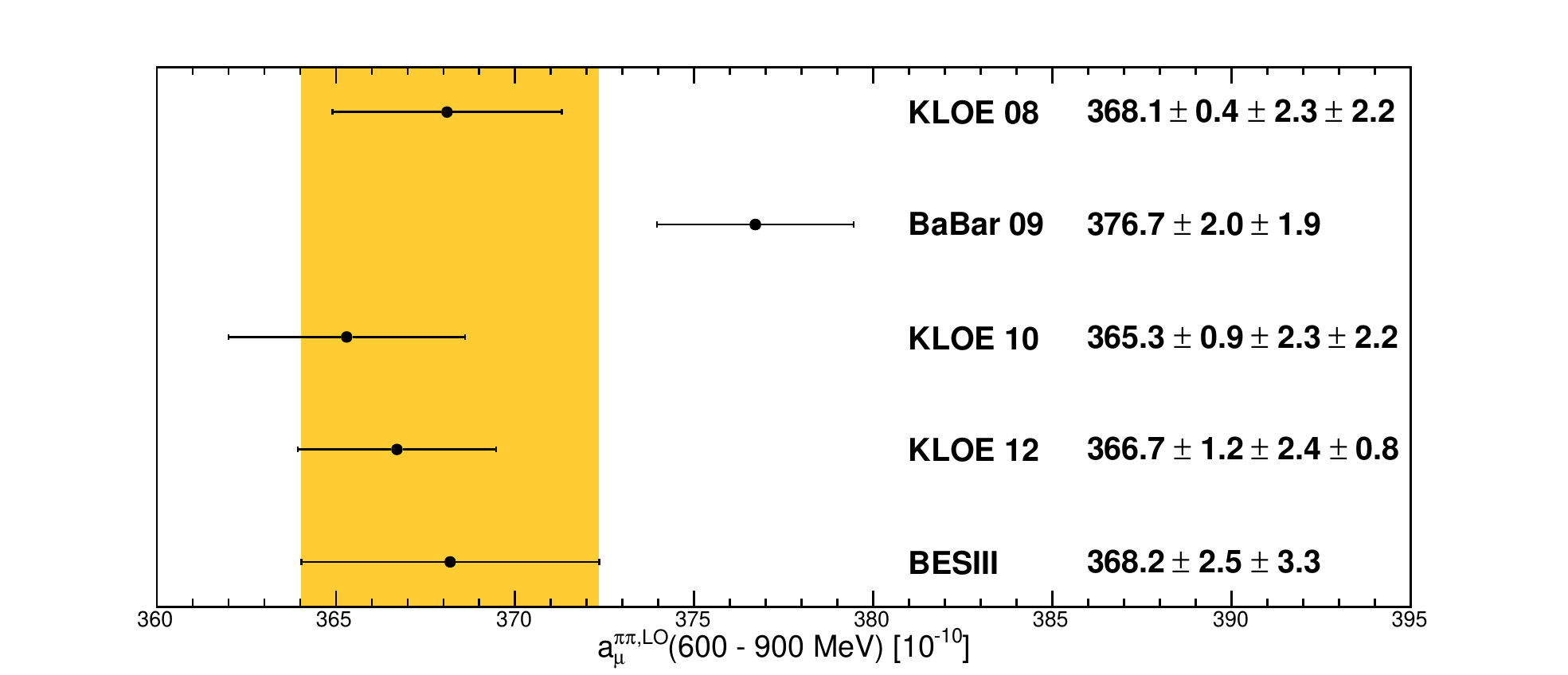}
 }
 \caption{\label{fig:pion} (From Ref.~\cite{Ablikim:2015orh}) \textbf{left:} Pion form factor determined at BESIII 
with the result of a Gounaris-Sakurai fit. \textbf{right:} Comparison of $a_\mu^{\pi\pi,LO}(600 - 900\,\textrm{MeV})$ 
determined at BESIII and by previous experiments.}
\end{figure}

The BESIII result is compared with previous measurements based on a fit of the pion form factor with a Gounaris-Sakurai 
parameterization, which is illustrated in Fig.~\ref{fig:pion}. Though systematic deviations can be observed for both the 
KLOE results as well as the BaBar results, the value of $a_\mu^{hVP,LO}$ obtained from the evaluation of the dispersive 
integral agrees nicely with the values obtained by the KLOE collaboration, as shown in Fig.~\ref{fig:pion}. The result 
of the BESIII measurement is $a_\mu^{\pi\pi,LO}(600 - 900\,\textrm{MeV}) = 368.2\pm2.5_{\rm stat}\pm3.3_{\rm 
syst}\cdot10^{-10}$. It confirms the deviation between the direct measurement and the SM prediction of $a_\mu$ to be on 
the level of more than three standard deviations. Recent dispersive evaluations of compilations of hadronic cross 
sections~\cite{Davier:2017zfy,Keshavarzi:2018mgv}, which include the BESIII result, were able to reduce the uncertainty 
of the hadronic vacuum contribution to $a_\mu$ by more than 20\%.

\section{\boldmath $e^+e^-\to\pi^+\pi^-\pi^0$}\label{sec:3pi}
The same data set of 2.93\,fb$^{-1}$ taken at $\sqrt{s}=3.773\,\textrm{GeV}$ is used to investigate the cross section of
$e^+e^-\to\pi^+\pi^-\pi^0$. In this analysis both the tagged and untagged ISR method are applied. The final result is 
obtained from the error weighted mean of both methods.

For the tagged analysis, events with two oppositely charged tracks and at least three photons are selected. A $5C$ 
kinematic fit, constrained by energy and momentum conservation, as well as the mass of the $\pi^0$ for two photon 
candidates, is applied to reject background and to settle the photon combinatorics. An additional requirement, where the 
assigned ISR photon combined with another photon candidate in the event shows an invariant mass close to the $\pi^0$ 
mass, can effectively suppress remaining background.

In the untagged analysis, events with at least two photon candidates are accepted. The kinematic fit is performed, 
considering the ISR photon an an unmeasured particle, with known mass but unknown momenta. Consequently, the number of 
constraints of the fit reduces to $2C$. Remaining background can be efficiently suppressed by rejecting all events where 
the photon four-momentum resulting from the kinematic fit is not pointing along the beam axes, i.e. 
$|\cos\theta_{\gamma_{\rm ISR}}|\leq0.9983$.

The dominating background contribution to both ISR methods comes from the (radiative) production of four pions in 
$e^+e^-\to\pi^+\pi^-\pi^0\pi^0(\gamma_{\rm ISR})$. It is studied in a separate analysis in order to tune MC simulations 
for reduced systematics due to background subtraction. Details of the analysis can be found in section~\ref{sec:4pi}.

\begin{figure}[htb]
 \centerline{
  \includegraphics[width=0.32\textwidth]{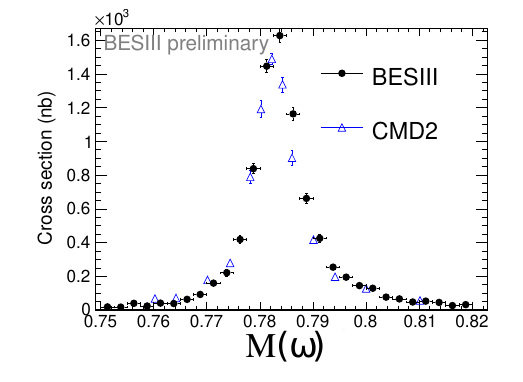}\hfill%
  \includegraphics[width=0.32\textwidth]{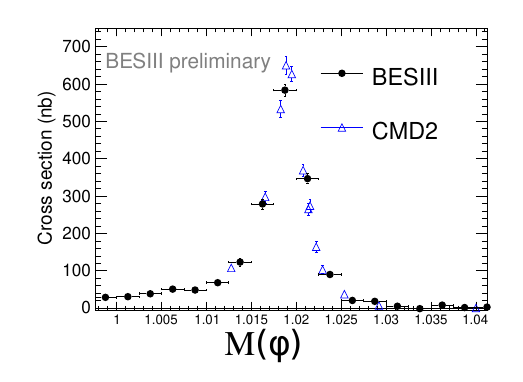}\hfill%
  \includegraphics[width=0.32\textwidth]{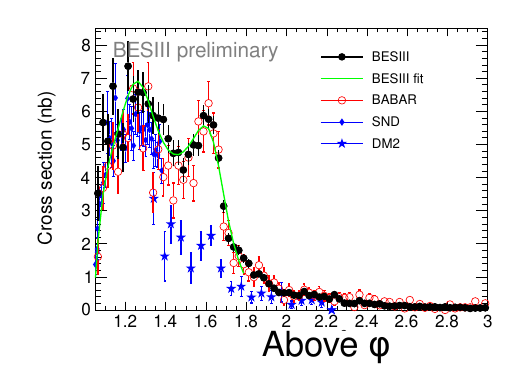}
 }
 \caption{\label{fig:3pi} Comparison of the preliminary BESIII results for the cross section of 
$e^+e^-\to\pi^+\pi^-\pi^0$ (solid black) and previous measurements. \textbf{left/center:} At the $\omega/\phi$ 
resonances with CMD-2 results~\cite{3pi_cmd2} (open blue). \textbf{right:} Above the $\phi$ resonances with 
Babar~\cite{3pi_babar} and DM2~\cite{3pi_dm2}. The green line shows a fit to the BESIII data.} 
\end{figure}

From the extracted signal event yields, the differential cross section is calculated. Figure~\ref{fig:3pi} shows the 
preliminary results for the cross section as function of $\sqrt{s}$. At the narrow resonances $\omega$ and $\phi$ the 
BESIII result can be compared with the results of the scan measurements performed at CMD-2~\cite{3pi_cmd2}. Good 
agreement is observed is both cases. The systematic uncertainty at the narrow resonances is better than 2\%. Above the 
$\phi(1020)$, the previous measurements are dominated BaBar result~\cite{3pi_babar}, obtained in an ISR measurement. 
Good agreement is observed with the BaBar result, including the structure attributed to the $\omega^{\prime\prime}$ 
resonance, which was not observed in the measurement by DM2~\cite{3pi_dm2}. A VMD inspired fit function can only 
describe the BESIII data by including this resonance. Furthermore, the preliminary result also allows to determine the 
branching ratio of $J/\psi\to\pi^+\pi^-\pi^0$. The final publication will also contain the resulting contribution of the 
three-pion channel to $a_\mu^{hVP,LO}$.

\section{\boldmath $e^+e^-\to\pi^+\pi^-\pi^0\pi^0$}\label{sec:4pi}
\begin{figure}[htb]
  \centerline{
  \includegraphics[trim=0mm 20cm 6cm 5mm, clip, width=0.48\textwidth]{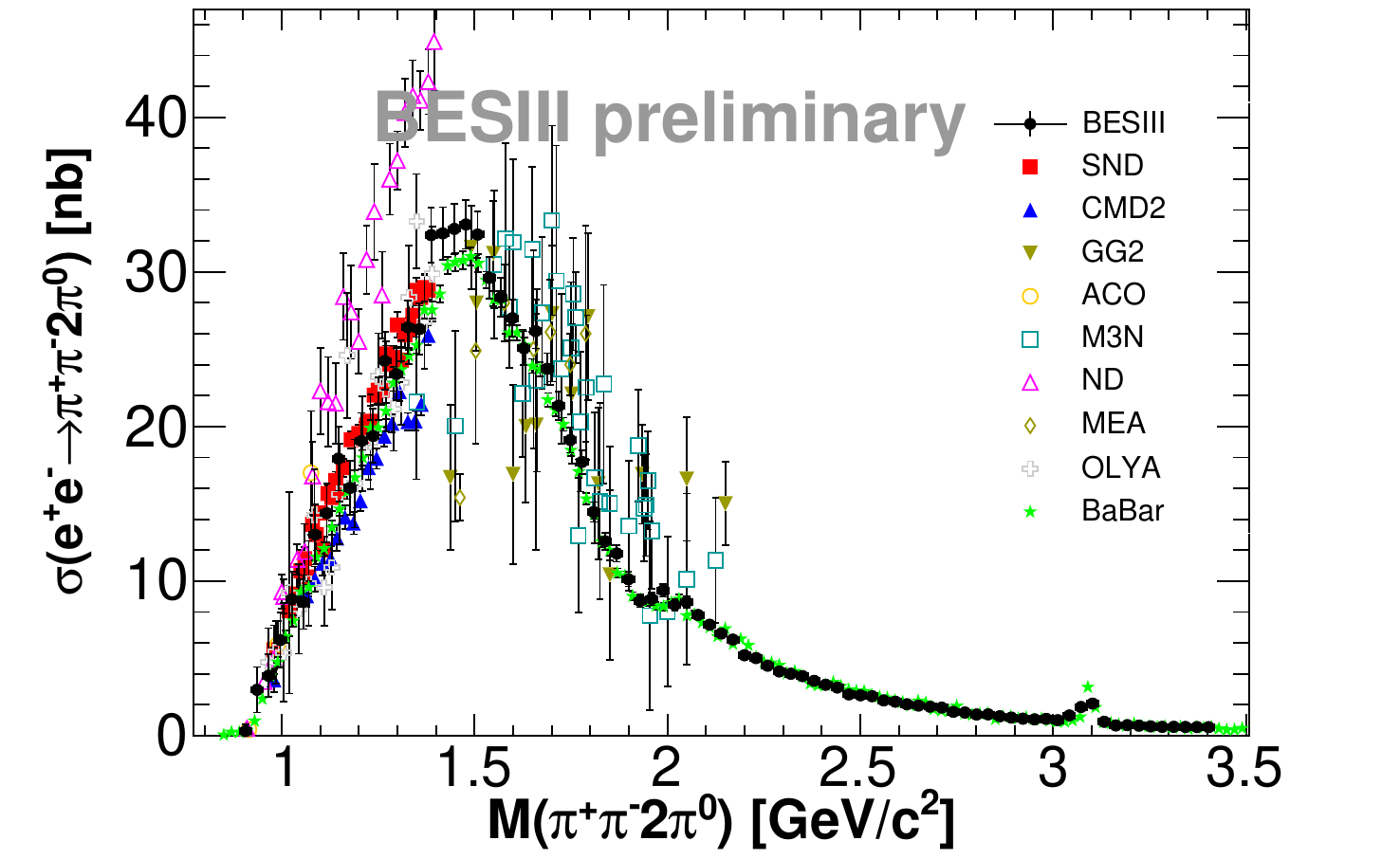}\hfill%
  \includegraphics[width=0.48\textwidth]{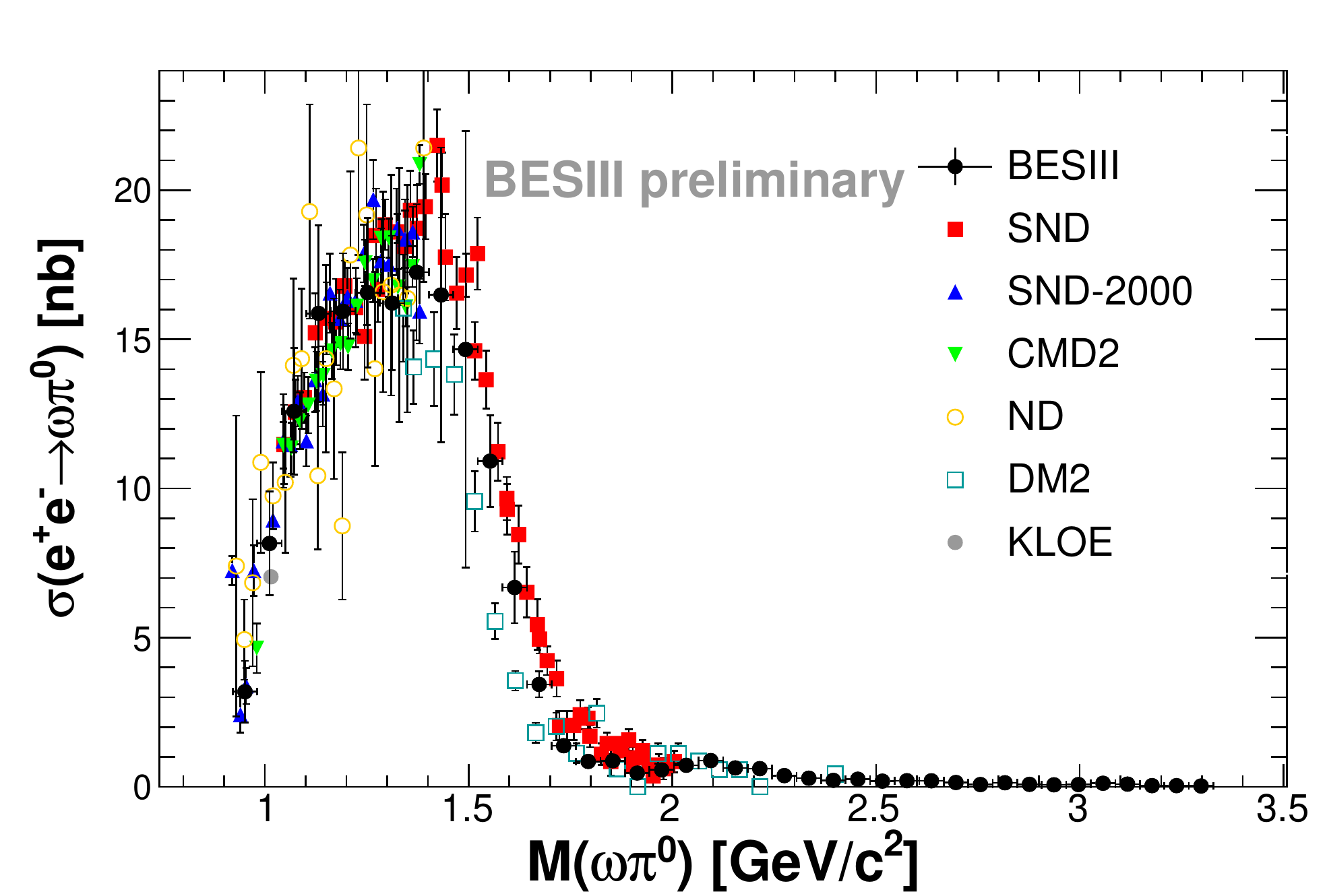}
 }
 \caption{\label{fig:4pi}\textbf{left:} The preliminary cross section of $e^+e^-\to\pi^+\pi^-\pi^0\pi^0$ from BESIII 
(solid black), compared to the BaBar result~\cite{BaBar:2017vzo} (solid green), and further previous scan measurements. 
\textbf{right:} The cross section of the sub-process $e^+e^-\to\omega\pi^0$ from BESIII (solid black) and previous 
scan experiments.}
\end{figure}

The analysis of the four pion final state is performed analogously to the strategy described in section~\ref{sec:3pi}, 
taking into account the higher photon multiplicity for the additional $\pi^0$. The tagged ISR method applies a $6C$ 
kinematic fit, exploiting the additionally possible constraint of another $\pi^0$ mass. In the same manner, a $3C$ fit 
is performed for the untagged ISR method. The dominating background contribution is the five-pion channel with three 
neutral pions. It is measured in a separate analysis in order to accommodate for a reliable background subtraction. 
Special attention is also paid to the systematic differences in the reconstruction efficiency of $\pi^0$. The total 
systematic uncertainty of the preliminary result is estimated to be on the level of 3\%. Figure~\ref{fig:4pi} shows the 
preliminary result of the cross section for $e^+e^-\to\pi^+\pi^-\pi^0\pi^0$. The BESIII result is in good agreement with 
the recently published BaBar result~\cite{BaBar:2017vzo}. Both results illustrate the potential of the ISR method to 
provide data with high accuracy over a wide energy range. The BESIII result has also been used to calculate the 
preliminary value of the contribution $a_\mu^{\pi^+\pi^.2\pi^0,LO}=18,63\pm0.27\pm0.57$, which agrees well within errors 
with the published value from BaBar.

Further investigations of the possible sub-processes in $e^+e^-\to\pi^+\pi^-\pi^0\pi^0$ have been carried out by 
studying the cross section as a function of the masses of different final state particle combinations. The important 
sub-process $e^+e^-\to\omega\pi^0$ is studied by determining the contribution of the $\omega$ resonance in the 
three-pion mass $M(\pi^+\pi^-\pi^0)$. The preliminary result of the cross section is shown in Fig.~\ref{fig:4pi}, and is 
in good agreement with previous measurements, while providing an improved accuracy.

\section{Outlook}
With the high accuracy measurement of the pion form factor, the BESIII collaboration provided already important input 
to the SM calculations of $a_\mu$. The accuracy of the current result is limited by systematics. An alternative 
approach, which normalizes the pion cross section to the muon yield, can reduce the uncertainties, but requires higher 
statistics. It is planned to extend the data set at $\sqrt{s}=3.773\,\textrm{GeV}$ to an integrated luminosity of 
$20\,\textrm{fb}^{-1}$. This data set will be large enough to determine the pion form factor with an expected accuracy  
on the level of 0.5\%.

Additionally, the hadronic cross sections at higher multiplicities are studied. The results for $e^+e^-\to \pi^+\pi^- 
\pi^0$ and $e^+e^-\to\pi^+\pi^-\pi^0\pi^0$ will be published in the near future and provide further important input for 
$a_\mu$. The cross section of $e^+e^-\to\pi^+\pi^-3\pi^0$, measured to determine the background contributions to 
$e^+e^-\to\pi^+\pi^-\pi^0\pi^0$ provides the first accurate measurement for the cross section in more than sixty years. 
So far, the channel has been taken into account in $a_\mu^{hVP}$ by evaluating isospin relations.

Apart from exclusive processes, also a measurement of the inclusive cross section ratio $R$ of hadron production to 
muon production is performed~\cite{Hu:2017lqp}. A scan over the full energy range covered by BEPCII has been performed, 
providing 130 scan points. The expected number of more than $10^5$ hadronic events at each energy allows for a 
measurement of the $R$ ratio, which is not limited by statistics. The final goal for the measurement is an accuracy of 
better than 3\%. It is expected that the dominating uncertainties come from the event generator 
\textsc{LundAreaLaw}~\cite{luarlw}, which is used to estimate the reconstruction efficiencies.

\end{document}